\documentclass[11pt,twoside]{article}
\usepackage{asp2010}

\resetcounters

\begin{document}

\title{Multiplicity of massive O stars and evolutionary implications}
\author{H. Sana,$^1$ S.E. de Mink,$^{2,3}$
A. de Koter,$^{1,4}$ N. Langer,$^5$
C.J. Evans,$^6$ M. Gieles,$^7$
E. Gosset,$^8$ R.G. Izzard,$^5$
J.-B. Le Bouquin,$^9$ F.R.N. Schneider$^5$
\affil{$^1$Astronomical Institute Anton Pannekoek, Amsterdam University, Netherlands}
\affil{$^2$Space Telescope Science Institute, Baltimore, USA}
\affil{$^3$Dept. of Physics and Astronomy, Johns Hopkins University, Baltimore, USA}
\affil{$^4$Astronomical Institute, Utrecht University, The Netherlands}
\affil{$^5$Argelander-Institut f\"ur Astronomie, Universit\"at Bonn, Germany}
\affil{$^6$UK Astronomy Technology Centre, Edinburgh, United Kingdom}
\affil{$^7$Institute of Astronomy, University of Cambridge, United Kingdom}
\affil{$^8$F.R.S.-FNRS, Institut d'Astrophysique, Li\`ege University, Belgium}
\affil{$^9$UJF-Grenoble 1 / CNRS-INSU, Institut de Plan\'etologie et d'Astrophysique de Grenoble, France}
}

\begin{abstract}
Nearby companions alter the evolution of massive stars in binary systems.  Using a sample of Galactic massive stars in nearby young clusters, we simultaneously measure all intrinsic binary characteristics relevant to quantify the frequency and nature of binary interactions. We find a large intrinsic binary fraction, a strong preference for short orbital periods and a flat distribution for the mass-ratios. Our results do not support the presence of a significant peak of equal-mass `twin' binaries. 
As a result of the measured distributions, we find that over seventy per cent of all massive stars exchange mass with a companion. Such a rate greatly exceeds previous estimates and implies that the majority of massive stars have their evolution strongly affected by interaction with a nearby companion.
\end{abstract}

Because stars expand as they evolve off the main sequence, massive stars in close binaries may exchange mass with their companion. The nature of a binary interaction is largely determined by the initial orbital period and mass ratio of the binary system. While it is recognized that the multiplicity rate of high mass stars is large \citep[e.g.,][]{MHG09,SaE11,CHN12}, the relative importance of binary- and single-star evolutionary scenarios remains poorly constrained because of uncertainties in our knowledge of intrinsic distributions of massive binary orbital parameters. This implies that predictions of the frequencies of supernova types, including hydrogen deficient core-collapse supernovae (CCSNe) and gamma ray bursts (GRBs) are also uncertain. We report here on our work to help alleviate these uncertainties; details of our analysis are given in \citet{SdMdK12}.

\section{Data sample}

In this investigation, we capitalize upon the large observational efforts in the past decades to spectroscopically observe and monitor massive O-type stars, and to characterize their multiplicity properties. We use information from over 1800 spectra of 71 O-type objects in six nearby young open clusters -- IC 1805 \citep{DBRM06,HGB06}, IC 1848 \citep{HGB06}, NGC 6231 \citep{SGN08}, NGC 6611 \citep{SGE09}, Tr 16 \citep[][and references therein]{RNF09} and IC 2944 \citep{SJG11}. The O stars in our sample have spectral types  from O9.7 to O3, corresponding to a mass range extending from 15 to 60 solar masses. 

With 40 identified spectroscopic systems, the observed binary fraction in our sample is $f_\mathrm{obs} = 0.56$. Combining new observations from VLT-UVES for long-period systems, and published results from detailed analysis of detected systems in the individual clusters, we consider the properties of these systems as a population. In total, 85\%\ and 78\%\ of our systems have constrained orbital periods and mass-ratios respectively. This allows us to derive the observed period and mass-ratio distributions of a statistically significant and homogeneous sample of massive stars.

The comparison between the present sample and the complete sample of Galactic O stars with multi-epoch spectroscopy is  discussed in \citet{SaE11}. Specifically the observed binary fractions from the two samples are identical  and the observed distributions of periods, mass-ratios and eccentricities  are compatible as revealed through a Kolmogorov-Smirnov (KS) test. We can thus assume that the multiplicity properties of our sample are representative of those of the Galactic O stars in general.

\begin{figure}[!t]
\plotone{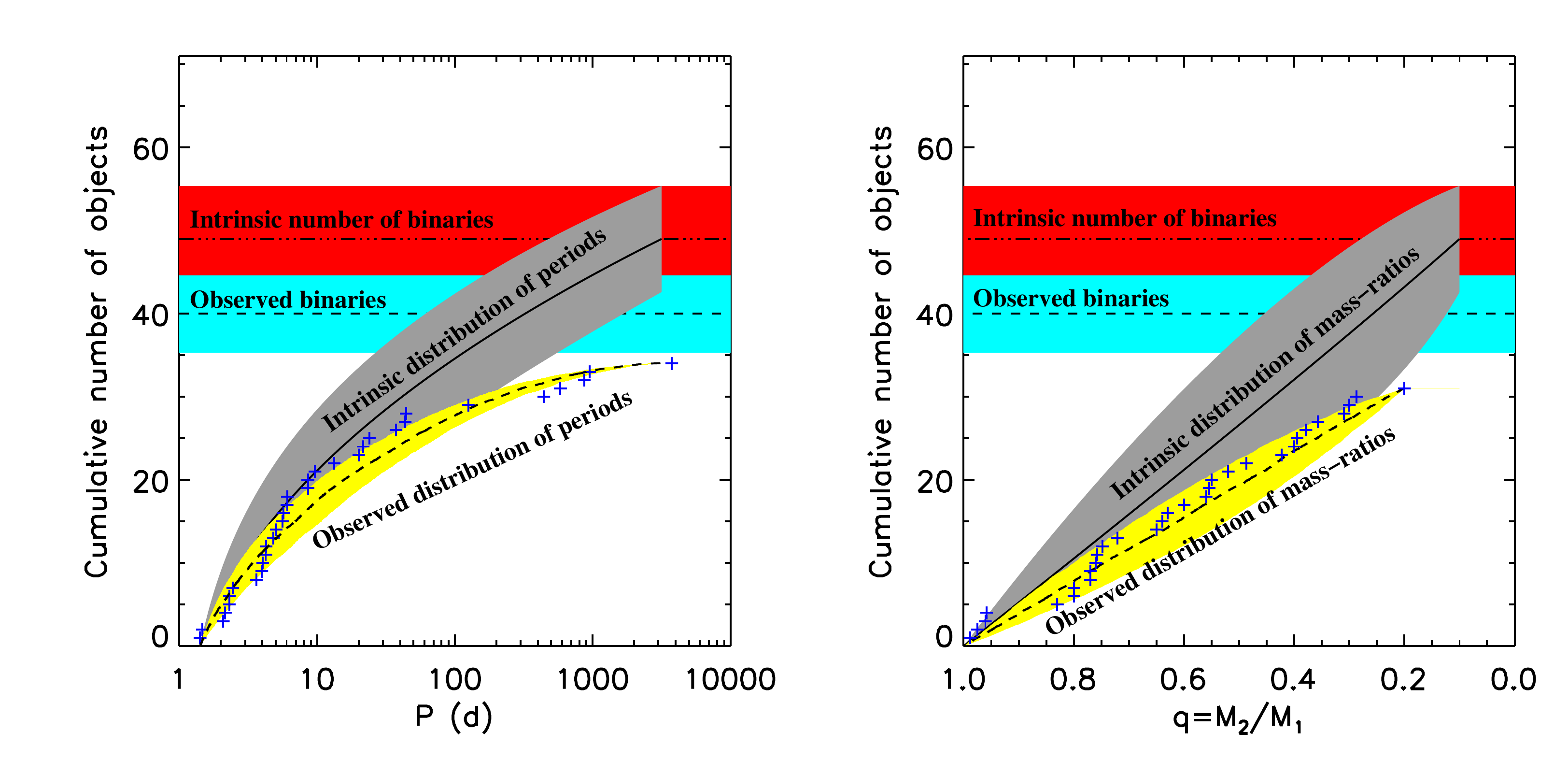}
\caption{Cumulative number distributions of orbital periods (left panel) and mass-ratios (right panel). Crosses give the observed distributions. Horizontal dashed and dashed-dotted lines indicate the observed and intrinsic number of binaries in our sample, respectively. Dashed and plain lines show, respectively, the best-fit simulated distributions to the data points and the corresponding intrinsic distributions. }\label{fig1}
\end{figure}

\section{Intrinsic multiplicity properties}
Observed distributions result from the convolution of the intrinsic distributions by the observational biases. We simulate observational biases using a Monte Carlo approach that incorporates the observational time series of each object in our sample, allowing us to estimate the fraction of undetected binaries and/or unconstrained orbital parameters. We adopt power laws for the probability density functions of orbital periods (in $\log_{10}$ space), mass-ratios and eccentricities. These power-law exponents and the intrinsic binary fraction are then simultaneously determined by a comparison of simulated populations of stars with our sample taking into account the observational biases. A more extensive discussion of similarities and differences between our approach and the methods of \citet{KoF07} and \citet{KiK12} is given in \citet{SdKdM12}.

We find an intrinsic binary fraction of $0.69 \pm 0.09$, a preference for close pairs ($f_\mathrm{\log P}\propto(\log P/\mathrm{d})^{-0.55}$) and a uniform distribution of the mass ratio ($f_\mathrm{q}\propto q^{-0.1}$) for binaries with periods up to about 3000 days. Figure 1 compares our intrinsic, simulated and observed cumulative distributions and shows that observational biases are mostly restricted to the longest periods and most extreme mass-ratios. We find no evidence of a preference for equal mass binaries. Compared to previous studies we obtain a steeper period distribution, thus a larger fraction of short period systems than previously thought. 

\begin{figure}[!t]
\plotone[width=4cm]{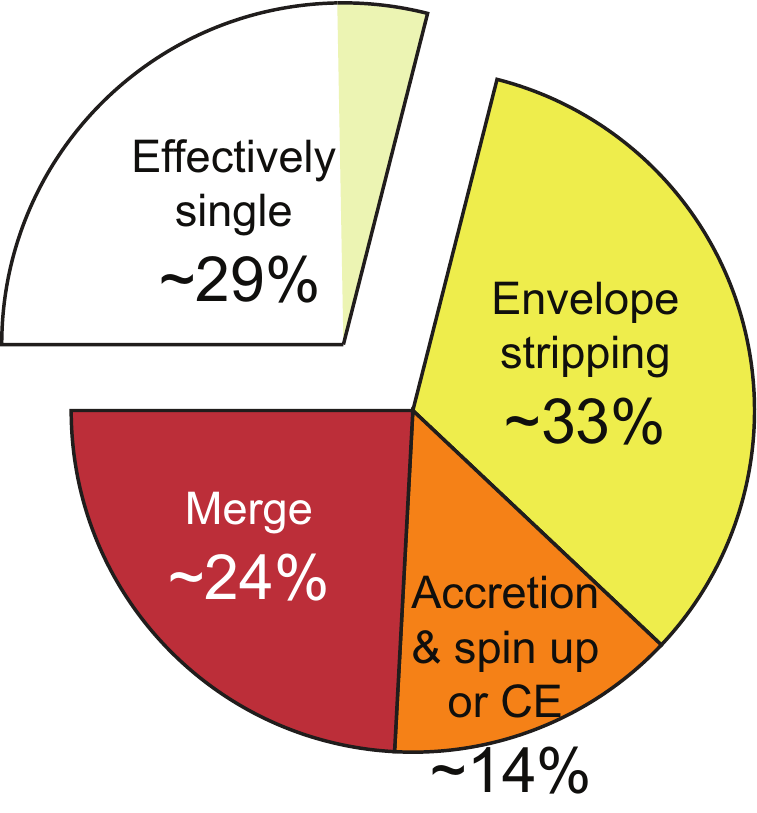}
\caption{Pie chart illustrating the rates of the various interaction scenarios. All numbers are expressed relative to the total number of stars born as O-type, including primaries, secondaries and single stars. }
\label{fig2}
\end{figure}

\section{Evolutionary implication}
 We determine the relative frequencies of binary interaction scenarios by integration of our distribution functions, under assumptions based on detailed  binary evolutionary calculations \citep{PJH92,Pol94,WLB01,dMPH07}. These assumptions are discussed in more details in \citet{SdMdK12}.

In short, the adopted upper limit for the orbital period at which mass-transfer interaction  plays a significant role is 1500 days. This approximately corresponds to maximum separation at which the initially most massive star will lose nearly all its entire hydrogen-rich envelope before the supernova explosion. The adopted orbital period upper limit for Case A Roche lobe overflow is 6 days. Systems with periods below 2 days are all expected to merge. Longer period Case A systems are assumed to merge if the mass-ratio is less than 0.65. A small fraction of mergers between an evolved star and a main sequence star (cases B and C mergers) are also expected but only affect 4\%\ of all O stars.

 Interactions that do not lead to coalescence are assumed to strip the primary stars. The secondary stars accrete material and angular momentum and are spun up to critical rotation velocities. Secondary stars which are O stars have been taken into account in our computation of the number of O stars affected by binary interaction, but lower mass (B-type) companions have been ignored. We further ignore lower mass stars that may become O stars during their life by gaining mass via mass accretion or via a binary merger. We also ignore triple systems. 

The  fraction of the various binary interaction channels are illustrated in Figure 2.
We find that $71 \pm 8$\%\ of all stars born as O stars are a member of a binary system that will interact by Roche lobe overflow. About 40\% of all O stars will be affected during their main sequence life-time, strongly impacting subsequent evolution. We predict that 33\%\ of O stars are stripped of their envelope before they  explode as hydrogen-deficient CCSNe (Types Ib, Ic and IIb). This fraction is remarkably close to the observed fraction of hydrogen-poor supernovae, i.e. 37\%\ of all CCSNe \citep{SLF11}. We also find that 20-30\%\ of all O stars will merge with a nearby companion.

 The interaction and merger rates that we compute are respectively two and three times larger than previous estimates  and result in a corresponding increase in the number of progenitors of key astrophysical objects produced by binary evolution such as double compact objects, hydrogen-deficient CCSNe and gamma-ray bursts.

\acknowledgements
Support for this work was provided by NASA through Hubble Fellowship grant HST-HF-51270.01-A awarded by the Space Telescope Science Institute, which is operated by the Association of Universities for Research in Astronomy, Inc., for NASA, under contract NAS 5-26555. SdM is Hubble fellow. 

\bibliography{/home/hsana/Desktop/Dropbox/Dropbox/literature}

\end{document}